\documentclass{epl}

\def\dj{\leavevmode\setbox0=\hbox{d}\dimen0=\wd0
        \setbox0=\hbox{d}\advance\dimen0 by -\wd0
        \rlap{d}\kern\dimen0\hbox to \wd0{\hss\accent'26}}
\def\DJ{\leavevmode\setbox0=\hbox{D}\dimen0=\wd0
        \setbox0=\hbox{D}\advance\dimen0 by -\wd0
        \rlap{D}\kern\dimen0\hbox to \wd0{\raise -0.4ex\hbox{\accent'26}\hss}}

\title{$^{8}$Li+$\alpha$ decay of $^{12}$B and its possible astrophysical implications}

\author{N. Soi\'{c}\inst{1} \and S. Cherubini\inst{2}  \and  M. Lattuada\inst{2}
 \and \DJ . Miljani\'{c}\inst{1}   \and  S. Romano\inst{2}  \and 
 C. Spitaleri\inst{2}  \and M. Zadro\inst{1} }

\institute{
  \inst{1} Ru\dj er Bo\v{s}kovi\'{c} Institute, Zagreb, Croatia \\
  \inst{2} INFN-Laboratori Nazionali del Sud and Universit\`{a} di Catania, Catania, Italy
}


\begin{document}

\maketitle

\shortauthor{N. Soi\'{c} \etal}

\begin{abstract}

 The $^{12}$B excitation energy spectrum has been obtained from coincidence
 measurements of the  $^{9}$Be+$^{7}$Li $\longrightarrow$ 2$\alpha$+$^{8}$Li 
 reaction at E$_{0}$=52 MeV. The decay of the states at excitations between 10
 and 16 Mev into $\alpha$+$^{8}$Li has been observed for the first time.
 Observed $\alpha$-decay indicates possible cluster structure of the $^{12}$B
 excited states.
 The influence of these states on the cross section of the astrophysically
 important $^{8}$Li($\alpha$,n)$^{11}$B and $^{9}$Be+t reactions is
 discussed and the results are compared with existing results.
\end{abstract}

\section{Introduction}

 Light nuclei are interesting in their own right as almost every nucleus
 possesses some unique properties which are fingerprints of its quantum
 structure. Their extremely varied structures, from spherical shell model
 shapes to prominent clustering, are a considerable challange to understand
 and model. Although light nuclei have been studied intensively for many decades,
 very limited and often contradictory information exist for many of them. That is
 also the case for  $^{12}$B nucleus. It is interesting to draw a comparison with
 its structure and structures of the $^{11}$Be, $^{11}$B, $^{12}$Be and $^{12}$C
 nuclei which all show prominent $\alpha$-clustering. As an example, $^{11}$B
 and $^{12}$C nuclei are typical cluster nuclei and their  $\alpha$-decays were
 experimentally observed in many particle coincidence measurements some 40 years ago
 (e.g. \cite{a1, a2}, for full list of the references see \cite{r10}). The helium
 decays of $^{12}$Be nucleus were reported only recently \cite{a3}, but there is no
 experimental results on the $\alpha$-decays of $^{12}$B. Some experimental
 indications for cluster structure of $^{12}$B states come from the measurements of the 
 $^{8}$Li($\alpha$,n)$^{11}$B reaction.
 According to inhomogenous models of big bang nucleosynthesis this
 reaction would have had a crucial role in
the production of A $\ge$ 12 nuclides \cite[\ and references therein]{r1}.
 Hence this reaction and
 the states of $^{12}$B at excitations close to the $\alpha$-$^{8}$Li
 separation energy (E$_{s}$=10.001 MeV) have attracted special interest.
 The 1996 status  of the results obtained from both direct and indirect
 measurements was critically reviewed in \cite{r3}. Results from a recent 
 direct measurement of the reaction \cite{r4} disagree with the
 conclusions from  \cite{r3}. The reaction cross section obtained
 in that exclusive measurement at c.m. energies between 1.5 and 7 MeV
 is almost a factor of 2 lower than that obtained in previous 
 inclusive measurements. Similar large discrepancies exist between two
 different theoretical approaches \cite{r5,r6}. With all the problems
 of direct measurements (radioactive  $^{8}$Li beam,  $^{4}$He gas target,
 detection of $^{11}$B ions and/or neutrons) one should not expect highly
 accurate data soon. In the meantime any new information about the states
 involved may help to narrow the uncertainties and may provide an additional
 quality check of the data obtained from direct measurements. Such an example,
 the $^{8}$Li+$\alpha$ decay of  $^{12}$B states having excitation energy,
 E$_{x}$, between 10 and 18 MeV, is presented here. 

\section{Results}

 The  $^{7}$Li+$^{9}$Be reaction measurements were performed 
 in order to study  $^{9}$Be and  $^{10}$Be nuclei \cite{r7,r8,r9}. The
 data also provide high quality results for the
 2$\alpha$+$^{8}$Li exit channel, which offer
 useful information on the  $^{12}$B states. The measurements were
 performed at the Laboratori Nazionali del Sud using a  $^{7}$Li$^{+++}$
 beam (E=52 MeV, I=60-100 nA) and a self-supported beryllium target
 (400 $\mu$g/cm$^{2}$). Outgoing charged particles were detected in
 detector telescopes consisting either of a very thin and a thick standard
 silicon surface barrier detector (T1), or of an ionization chamber and a silicon 
 position sensitive detector (T2). The angular range covered by the T2 telescope
 was 8$^{\circ}$, while the T1 angular opening was 1.5$^{\circ}$. 
 Three T1 telescopes were positioned
 on one side and two T2 on the other side of the beam, all in the same
 scattering plane. Coincident events of any T1-T2 pair were recorded.
 The experimental details were published elsewhere \cite{r7,r8,r9}.

\begin{figure}
 \twofigures[scale=0.4]{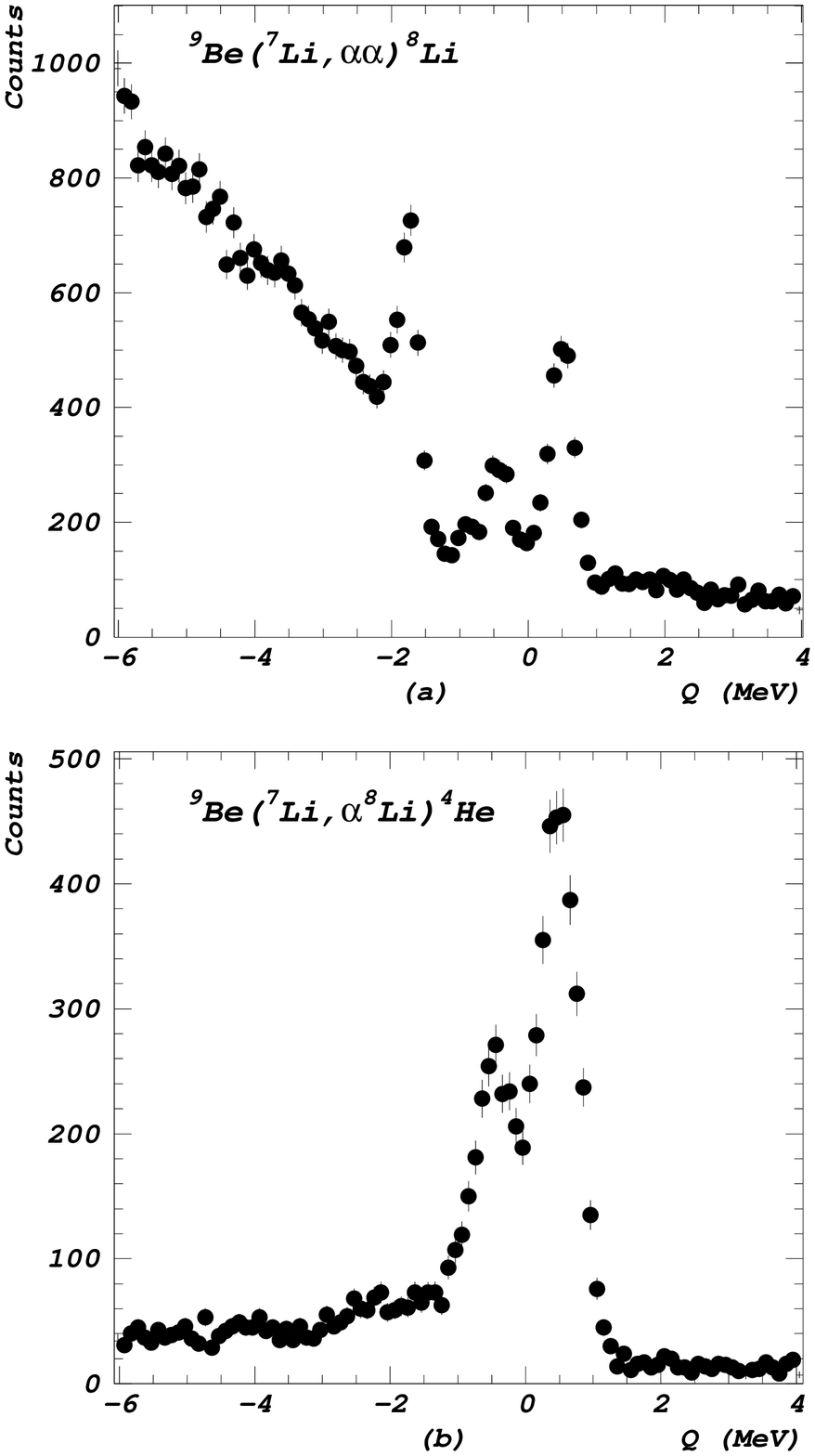}{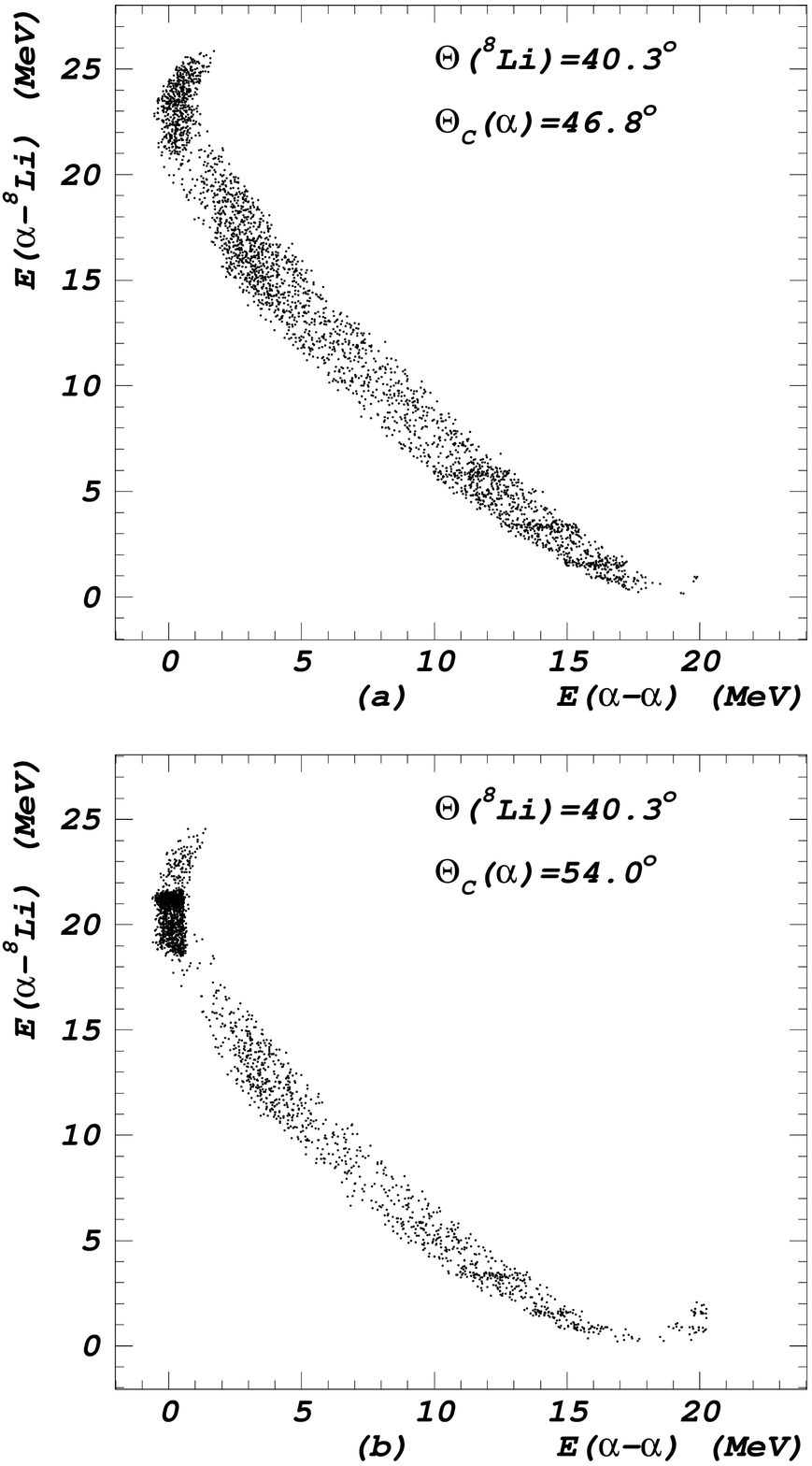}
 \caption{Typical examples of the Q-value spectra reconstructed from 
 measured $\alpha$-$\alpha$ (a) and $\alpha$-$^{8}$Li (b) coincident
 events.  The highest energy peak corresponds to the $^{8}$Li ground
 state and lower are $^{8}$Li excitations.}
\label{q}
\caption{ Typical examples of the $\alpha$-$^{8}$Li coincident
 data for two detectors settings presented in 2D plots of relative energies
 E($\alpha$-$^{8}$Li) vs  E($\alpha$-$\alpha$). Strips of data points parallel to the axes
 correspond to resonances in respective two-body systems. The most prominent 
 $^{12}$B states are at the excitation energies of 10.9, 11.6, 13.3 and 15.7 MeV. 
 The ground and
 first excited state of $^{8}$Be are also populated, but don't overlap with
  $^{12}$B states below 20 MeV in excitation. }
\label{erer}
 \end{figure}

 The measurement of the energies and angles of detected particles permitted
 the kinematics of the $^{7}$Li+$^{9}$Be $\longrightarrow$ 2$\alpha$+$^{8}$Li
 reaction (Q=0.460 MeV) to be fully reconstructed. Figure \ref{q} shows the
 Q-value spectra for that reaction for particular T1-T2 pair in the case
 of $\alpha$-$\alpha$
 (a) and $\alpha$-$^{8}$Li (b) coincidences. In both spectra the highest
 energy peak corresponds to the $^{8}$Li ground state (J$^\pi$=2$^+$) and
 the next one
 is the 0.981 MeV, 1$^+$ state. For $\alpha$-$\alpha$ events there is also 
 another peak which corresponds to the unbound $^{8}$Li second excited state
 (3$^+$) at 2.255 MeV. The obtained resolution, 500 keV for $\alpha$-$\alpha$ and
 800 keV for  $\alpha$-$^{8}$Li events, is sufficient to resolve these
 $^{8}$Li states. These resolutions are worse than the particle energy
 resolution (150-250 keV for $\alpha$ and 300-400 keV for $^{8}$Li)
 because all the uncertainties in the energies and angles of detected
 particles contribute to the error in determination of the total energy.
 Inferior resolution in the latter case is due to the light mass of the recoil particle,
 because a small uncertainty in the measured momenta gives a larger uncertainty
 in the energy of the unobserved particle.
 Part of the background events in both spectra is due to the
 $^{7}$Li+$\alpha$ coincidences leaking through the $^{8}$Li selection windows.
 Main background contributions for the $\alpha$-$\alpha$ events come from the
 reactions on carbon and oxygen impurities in the target, which are not
 present in the case of $\alpha$-$^{8}$Li events due to the very negative
 Q-values for the ($^{7}$Li,$\alpha$$^{8}$Li) reaction on these nuclei.

 Fig. \ref{erer} presents two examples of collected data in 2-D
 E($\alpha$-$\alpha$) vs. E($\alpha$-$^{8}$Li) plots of coincident 
 $\alpha$-$^{8}$Li events. E(x-y) are the c.m. energies in 
 respective two-body systems obtained directly from the measured energies
 and angles of $^{8}$Li and $\alpha$. Excitation energy in the
 systems is given by E$_{x}$=E(x-y)+E$_{s}$; in the $^{12}$B
 case that is  E$_{x}$($^{12}$B)=E($\alpha$-$^{8}$Li)+10.001 MeV.
 These spectra were obtained by selecting only the events associated
 with the 2$\alpha$+$^{8}$Li(gs) final state. In both cases  $^{8}$Li
 were detected in a T1 telescope at 40$^{\circ}$ together with  
 $\alpha$-particles detected in a telescope T2 once set at an angle of
 46.8$^{\circ}$ and other time at 54$^{\circ}$. One can notice
 horizontal and vertical groupings of the events corresponding to
 resonances in the systems i.e. to states in $^{8}$Be  and
 $^{12}$B, respectively. The excitation energies of observed
  $^{8}$Be ground and first excited state are well reproduced
 and their contributions are well separated from the contributions
 of the  $^{12}$B states below 20 MeV in excitation.

 \begin{figure}
 \onefigure[scale=0.65]{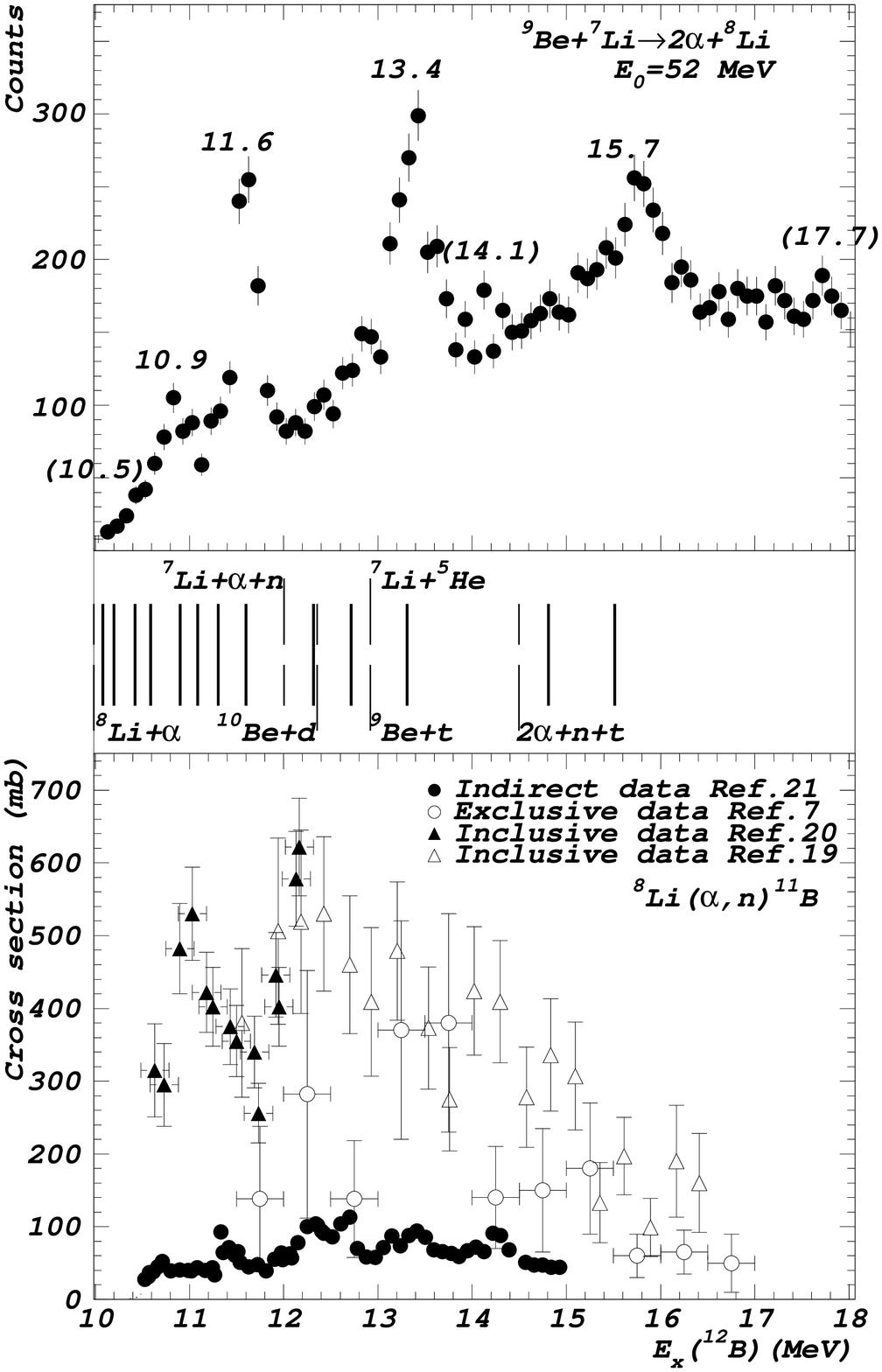}
 \caption{ The $^{12}$B excitation energy spectrum from the present
 $^{7}$Li+$^{9}$Be $\longrightarrow$ 2$\alpha$+$^{8}$Li reaction
 measurements (top); adopted $^{12}$B level scheme for excitations
 between 10 and 18 MeV \cite{r10} (center); and
 excitation functions from the measurement of $^{11}$B(n,$^{8}$Li)$^{4}$He
 reaction \cite{r19}, and measurements of the $^{4}$He($^{8}$Li,$^{11}$B)n
 reaction \cite{r4, r17, r18}(bottom). Observed peaks in the present spectrum
 are labelled with excitation energies (MeV).
 Known levels are marked with bold solid lines and the decay thresholds of the
 labelled channels with dashed lines. }
 \label{tot}
 \end{figure}

 On Fig. \ref{tot} are displayed together: 
 a $^{12}$B excitation energy spectrum from the present $^{7}$Li+$^{9}$Be
 $\longrightarrow$ 2$\alpha$+$^{8}$Li reaction measurements (top),
 $^{12}$B level scheme (center) and
 previously measured excitation functions of $^{4}$He($^{8}$Li,$^{11}$B)n
 and $^{11}$B(n,$^{8}$Li)$^{4}$He reactions (bottom).
 The level scheme for excitations between 10 and 18 MeV is adopted from \cite{r10}. 
 It was obtained mainly from the older measurements of the
 $^{9}$Be($^{7}$Li,$\alpha$) and $^{10}$B(t,p) reactions \cite{r11,r12,r13}
 and, for 10 $<$ E$_{x}$ $<$ 13.5 MeV, partially confirmed in more
 recent measurements of the  $^{9}$Be($\alpha$,p) and $^{11}$B(d,p)
 reactions \cite{r14, r15, r16}. Some of decay thresholds are also indicated.
 The most relevant one here is that for $\alpha$+$^{8}$Li decay at
 10.001 MeV. The threshold for n+$^{11}$B decay is at 3.37 MeV, which
 means that, for E$_{x}$ $>$ 10 MeV, channels for neutron decay 
 into high excited states of $^{11}$B are also opening.
 Similarly, some channels corresponding to excited states of other 
 residual nuclei ($^{8}$Li$^{*}$, $^{7}$Li$^{*}$ etc.) are also open
 at these energies. 
 Having all this in mind, as well as the fact that there are 18
 $^{12}$B levels for 7 $<$ E$_{x}$ $<$ 10 MeV, the level scheme for 
  10 $<$ E$_{x}$ $<$ 18 MeV reflects more accurately our insufficient knowledge
 rather than reality.

  Excitation function data for the  $^{4}$He($^{8}$Li,$^{11}$B)n reaction
 \cite{r4, r17, r18}, shown in Fig. \ref{tot} (bottom), represent the
 sums of transitions involving $^{8}$Li ground state and different
 $^{11}$B states. In all of the cases of direct measurements $^{11}$B
 ions were detected and only in one measurement \cite{r4} neutrons were also 
detected. As can be seen, the quality of the data, as well as the agreement 
 between the inclusive \cite{r17,r18} and exclusive measurements \cite{r4}
 are not good. Also, only three data points fall into the upper part of the 
 most interesting energy region (10-11 MeV) for big bang nucleosynthesis
 analyses. The data on the $^{11}$B(n,$^{8}$Li)$^{4}$He reaction \cite{r19}
 were obtained by the detection of 2$\alpha$ following $\beta^{-}$-decay
 of $^{8}$Li. Up to E$_{x}$=11 MeV the only contribution comes from the 
 $^{11}$B(n,$\alpha_{0}$)$^{8}$Li reaction, above this energy the reaction
 leading to the $^{8}$Li first excited state could also contribute. From similar
 measurements more refined data for 10.3 $<$ E$_{x}$ $<$ 11 MeV were also
 obtained \cite{r20}.

 The $^{12}$B excitation energy spectrum (Fig. \ref{tot} top)
 was obtained from spectra shown on Fig. \ref{erer} by projecting
 it to the E($\alpha$-$^{8}$Li) axis. It represents a sum of coincident
 events recorded by all the T1-T2 telescope pairs, corresponding to the
 exit channel of two $\alpha$-particles and the $^{8}$Li ground state. It 
 should be mentioned that the data were collected at large $\alpha$-particle
 center-of-mass angles ($>$ 35$^{\circ}$), where both direct reaction mechanisms,
 triton stripping off $^{7}$Li and  $^{5}$He pick-up from $^{9}$Be, may
 contribute to the $^{9}$Be($^{7}$Li,$\alpha$)$^{12}$B reaction. A part
 of the smooth ''background'' in the $^{12}$B spectrum is due to the 
  $^{9}$Be($^{7}$Li,$^{8}$Li)$^{8}$Be reaction to different 2$\alpha$
 decaying states of $^{8}$Be. From the spectrum it is evident that the
 strongest contributions in these kinematical conditions come from the
 states at 10.9, 11.6, 13.4 and 15.7 MeV, with weaker or no contributions from
 other known states of $^{12}$B. Error in excitation energy for these
 states, determined from the excitation energies of the known $\alpha$-decaying states
 of the $^{10}$B and $^{11}$B nuclei, is $\leq$100 keV and resolution,
 estimated from the width of the 10.9 and 11.6 MeV states, is $\sim$300 keV.
  Another interesting result of the present
 measurement is that our data show no evidence for $\alpha$-decay of
 $^{12}$B states to the first two excited states of $^{8}$Li. 
 That is probably consequence of the kinematical conditions of these measurements.
 The main contributions to these exit channels come from one neutron transfer
 reactions forming excited states of $^{8}$Li and $^{8}$Be, which mask contributions
 from the $^{12}$B states.

\section{Discussion}

 It should be mentioned that the excitation spectrum
 from the present measurements emphasizes $^{12}$B states with large $\alpha$
 and t or $^{5}$He spectroscopic factors, i.e. states with well
 developed cluster structure, whereas resonant contributions to the ($^{8}$Li,n)
 reaction come from $^{12}$B states with significant both  $\alpha$ and
 neutron partial widths. However, it is resonable to expect neutron 
 decay of all states at these high excitation energies, thus all observed
 $\alpha$-decaying states may influence the reaction rate. Weak neutron
 decay for some of these $^{12}$B states would indicate its exotic cluster structure.

 The most important energy region (10-11 MeV) for the analysis of the 
 $^{8}$Li($\alpha$,n)$^{11}$B reaction role in big bang nucleosynthesis
 was only partially covered in present measurements due mainly to
 kinematical conditions of the experiment, thus detection efficiency for that 
 energy region decreases rapidly (its estimated value for E$_x$=11.0 MeV
 is a factor of 2.5 and for E$_x$=10.6 MeV a factor of 7 lower than
 for E$_x$=11.5 MeV). Very weak contributions of the state(s) between 10.4
 and 10.6 MeV may be present in some individual spectra, but no definite
 conclusions can be made.

 The presence of the 10.9 MeV state is evident in almost all of the individual spectra,
 the sum spectrum, as well as in the spectra from the ($^{7}$Li,$\alpha$),
 ($\alpha$,p), (t,p) and (d,p) reactions \cite{r11, r13,r14,r15, r16}. This state
 appears as a very weak peak in the (n,$^{8}$Li) reaction excitation function
 or in the ($\alpha$,n$_{0}$) S-factor \cite{r3, r19, r20} and noticeably
 stronger in the data of one of inclusive ($^{8}$Li,$^{11}$B) reaction
 measurements \cite{r18}. From the neutron decay data \cite{r16} it was
 concluded that it is a 3$^{+}$ state and that its $\Gamma_{\alpha}$ =
 1/5 $\Gamma_{tot}$. On the other hand, Descouvemont \cite{r6} puts the 3$^{+}$ state
 somewhat higher (12.0 MeV) and with a much larger $\Gamma_{\alpha}$ 
 ($\Gamma_{\alpha}$ = 2/3 $\Gamma_{tot}$) reflecting its $^{8}$Li-$\alpha$
 ``molecular'' structure. The present result shows its conspicuous $\alpha$-decay
 indicating that the 10.9 MeV state could contribute significantly to the 
 ($^{8}$Li,$^{11}$B) reaction cross section.

  The states at 11.3 and 11.6 MeV are both seen in the ($^{7}$Li,$\alpha$) 
 and ($\alpha$,p) reactions, in all the cases the higher state was more
 strongly populated.
 The present result shows almost exclusive population of the 11.6 MeV state, which 
 probably reflects not only its preferential feeding by the primary reaction but also
 its larger $^{8}$Li-$\alpha$ decay probability. The data on the (n,$^{8}$Li)
 reaction indicate that the lower state is more strongly excited in the process.
 The prominent $\alpha$-decay of the 11.6 MeV state, its strong population
 in triton transfer reactions and its very weak contribution in neutron transfer
 reactions and also in the $^{4}$He($^{8}$Li,n) reaction, 
 suggest that this state has a small neutron partial width and may
 possess exotic $\alpha$-cluster structure.

 Very strong population of the state(s) around 13.3 MeV is observed here. It was
 claimed that the state at 13.3 MeV has a width of 50 \cite{r11} or 55 keV \cite{r14}.
 A much broader structure around that energy was observed in the (t,p) \cite{r13}
 and also in the ($\gamma$,$\pi^{+}$) reaction \cite{r21}.
 The resolution and statistics of present results can not exclude the possibility
 that a broader state or several states are involved. The $^{9}$Be($^{7}$Li,$\alpha$)
 reaction to the state has the largest cross section at forward angles
 ($<$ 40$^{\circ}$) of all transitions to $^{12}$B states \cite{r12}. This was 
 attributed to the so called ``threshold'' state having the t+$^{9}$Be cluster
 structure according to theoretical arguments by Baz \cite{r22} (later elaborated 
 also by others \cite{r24} and experimentally confirmed for many nuclei). 
 The structure in the (n,$^{8}$Li) excitation function \cite{r19} as well as the
 maximum cross section observed at these energies in the most recent
 ($^{8}$Li,$^{11}$B) reaction measurement \cite{r4} may also be attributed to this
 state. Being only a few hundred keV above the t+$^{9}$Be reaction 
 threshold (i.e. inside the Gamow peak), it would have a large influence 
 on the thermonuclear $^{9}$Be+t reaction rates. Boyd at al \cite{r14} discussed
 the possible importance of the $^{9}$Be(t,n)$^{11}$B reaction in the production of
 $^{11}$B in primordial nucleosynthesis. They calculated its rate under the assumption 
 that the sum of the neutron partial widths to all $^{11}$B states is very nearly
 equal to the total width, which does not seem to be supported by
 the present result of a strong  $\alpha$-decay.
 One can expect that the states with $^{9}$Be+t or $^{7}$Li+$^{5}$He (or better 
 $\alpha$+$^{5}$He+t) clustering below the Coulomb barrier for these channels,
 have relatively large $\alpha$+$^{8}$Li$_{gs}$ decay width. Also at these energies,
 in addition to n+$^{11}$B and $\alpha$+$^{8}$Li$_{gs}$, decays to the first
 two  $^{8}$Li excited states as well as to $^{10}$Be+d are possible and their
 widths may not be negligible. Obviously, the thermonuclear
 $^{9}$Be(t,$\alpha$)$^{8}$Li reaction rate should be strongly affected by this state,
 claimed to be a 2$^{+}$ state \cite{r11, r14}.

 A broad peak at 15.7 MeV may correspond to a state claimed to be at 15.5 MeV in the
 ($^{7}$Li,$\alpha$) reaction spectrum measured at E$_{0}$=30 MeV and
 $\Theta$=0$^{\circ}$ \cite{r12}, which is the only measurement reporting state(s)
 at these excitations. It should be mentioned that close to these energies are the
 thresholds for the following
 decays: $^{5}$He+t+$\alpha$ (15.39 MeV) and  $\alpha$+$^{8}$Li$^{*}$ (4$^{th}$
 excited state) (15.4 MeV) and it may be one of those threshold cluster states.
 At higher energies there is no evidence for other peaks except maybe at 17.7 MeV.

\section{Conclusion}

 The $\alpha$+$^{8}$Li decay of several  $^{12}$B
 states for 10 $<$E$_{x}$$<$ 16 MeV has been observed for the first time. The most
 strongly populated states are at 10.9, 11.6, 13.3 and 15.7 MeV, which is the
 consequence of their cluster structure. The 11.6 MeV state is a good candidate 
 for having an exotic cluster structure. Some of these states may have a strong influence 
 on the cross section of the $^{8}$Li($\alpha$,n)$^{11}$B reaction which is, in
 part, supported by the (n,$^{8}$Li) reaction results and also by the crude data from
 the ($^{8}$Li,$^{11}$B) reaction measurements. From the present results one can also
 expect that the thermonuclear $^{9}$Be+t reaction rates are strongly
 influenced by the 13.3 MeV $\alpha$-decaying state(s), for which the excitation energy 
 is only 0.4 MeV above the t+$^{9}$Be threshold. The present results also show
 that the indirect method of coincidence measurement
 of many-body nuclear reactions can give important information for key astrophysical
 reactions. Because one cannot soon expect more accurate data on the astrophysically
 important $^{8}$Li($\alpha$,n)$^{11}$B reaction, it would be much easier to more
 thoroughly explore the excitation energy region above 10 MeV in 
 experiments similar to the present one, with reactions like
 $^{7}$Li($^{6}$Li,p), $^{7}$Li($^{7}$Li,d), $^{9}$Be($\alpha$,p),
  $^{11}$B(d,p), $^{13}$C(d,$^{3}$He),$^{14}$C(d,$\alpha$) etc. In these measurements
 one may simultaneously determine E$_{x}$, $\Gamma_{tot}$, $\Gamma_{\alpha}$,
 $\Gamma_{n}$, and in some cases even the spin, of all of the states in the region,
 which will make it possible to
 determine the resonant contributions, the dominant part of the reaction 
 cross section at these low energies.

\end{document}